\documentclass[acmsmall,screen,nonacm]{acmart}

\usepackage{subcaption}

\AtBeginDocument{%
  \providecommand\BibTeX{{%
    \normalfont B\kern-0.5em{\scshape i\kern-0.25em b}\kern-0.8em\TeX}}}

\setcopyright{acmcopyright}
\copyrightyear{2021}

\acmConference[What Can CHI Do About Dark Patterns?]{Position Paper at the Workshop "What Can CHI Do About Dark Patterns?" at the CHI Conference on Human Factors in Computing Systems}{May 8--13, 2021}{Online Virtual Conference (originally Yokohama, Japan)}

\makeatletter
\renewcommand{\@copyrightowner}{Copyright for this paper by its authors. Use permitted under Creative Commons License Attribution 4.0 International (CC BY 4.0}
\makeatother

\begin{document}

\title{All in one stroke? Intervention Spaces for Dark Patterns}

 \author{Arianna Rossi}
\email{arianna.rossi@uni.lu}
\affiliation{
  \institution{SnT, University of Luxembourg}
  \country{Luxembourg}
  \city{Luxembourg}
  }
\author{Kerstin Bongard-Blanchy}
\email{kerstin.bongard-blanchy@uni.lu}
\affiliation{
  \institution{University of Luxembourg}
  \country{Luxembourg}
   \city{Esch-sur-Alzette}
  }

\renewcommand{\shortauthors}{Rossi and Bongard-Blanchy, 2021}

\begin{abstract}
This position paper draws from the complexity of dark patterns to develop arguments for differentiated interventions. We propose a matrix of interventions with a \textit{measure axis} (from user-directed to environment-directed) and a \textit{scope axis} (from general to specific). We furthermore discuss a set of interventions situated in different fields of the intervention spaces. The discussions at the 2021 CHI workshop "What can CHI do about dark patterns?" should help hone the matrix structure and fill its fields with specific intervention proposals.
\newline
    \begin{flushleft}
    Arianna Rossi and Kerstin Bongard-Blanchy. 2021. All in one Stroke? Intervention Spaces for Dark Patterns. \textit{Position Paper at the Workshop "What Can CHI Do About Dark Patterns?" at the CHI Conference on Human Factors in Computing Systems (CHI'21), May 8--13, 2021, Online Virtual Conference (originally Yokohama, Japan).} 5 pages. https://arxiv.org/abs/2103.08483
    \end{flushleft}
\end{abstract}

\maketitle
\section{One definition for a multifaceted phenomenon}
Building on \citet{mathur2019dark} and \citet{thaler2008nudge}, we propose the following definition: \textbf{Dark patterns are interface designs that materialise manipulation strategies to benefit a service. The designs seek to nudge, steer, deceive or coerce users into taking actions or decisions they might not have taken if they had paid full attention and possessed complete information, unlimited cognitive abilities, and complete self-control}. Dark patterns vary in \textbf{design elements} (e.g., visuals, text), \textbf{design attributes} (e.g., asymmetric, covert, deceptive), and their \textbf{impact} (on individual or public welfare, regulatory objectives, and user autonomy) \cite{Mathur2021what}. Differences also appear on the user perspective: individuals might find certain instances of dark patterns more or less \textbf{recognisable}, \textbf{resistible}, and \textbf{acceptable} than others. 

\paragraph{Recognisable yet irresistible, but sometimes acceptable} The above-cited differences from the user perspective emerged in a survey (n=406) that we conducted to assess people's awareness of dark patterns \cite{bongard2021}. We found that respondents are \textbf{aware of their exposure to manipulative designs} and that many, especially the generations who grew up with the internet, are \textbf{able to recognise them}. However, most individuals \textbf{cannot precisely determine the consequences} of dark pattern influence. They thus display little concern. Moreover, participants' \textbf{likelihood of being influenced does not correlate with their general awareness}, even though those who are able to recognise a more significant number of dark patterns estimate their likelihood of being influenced lower.

\begin{figure}[btp]
    \begin{subfigure}{0.49\textwidth}
        \includegraphics[width=0.9\linewidth]{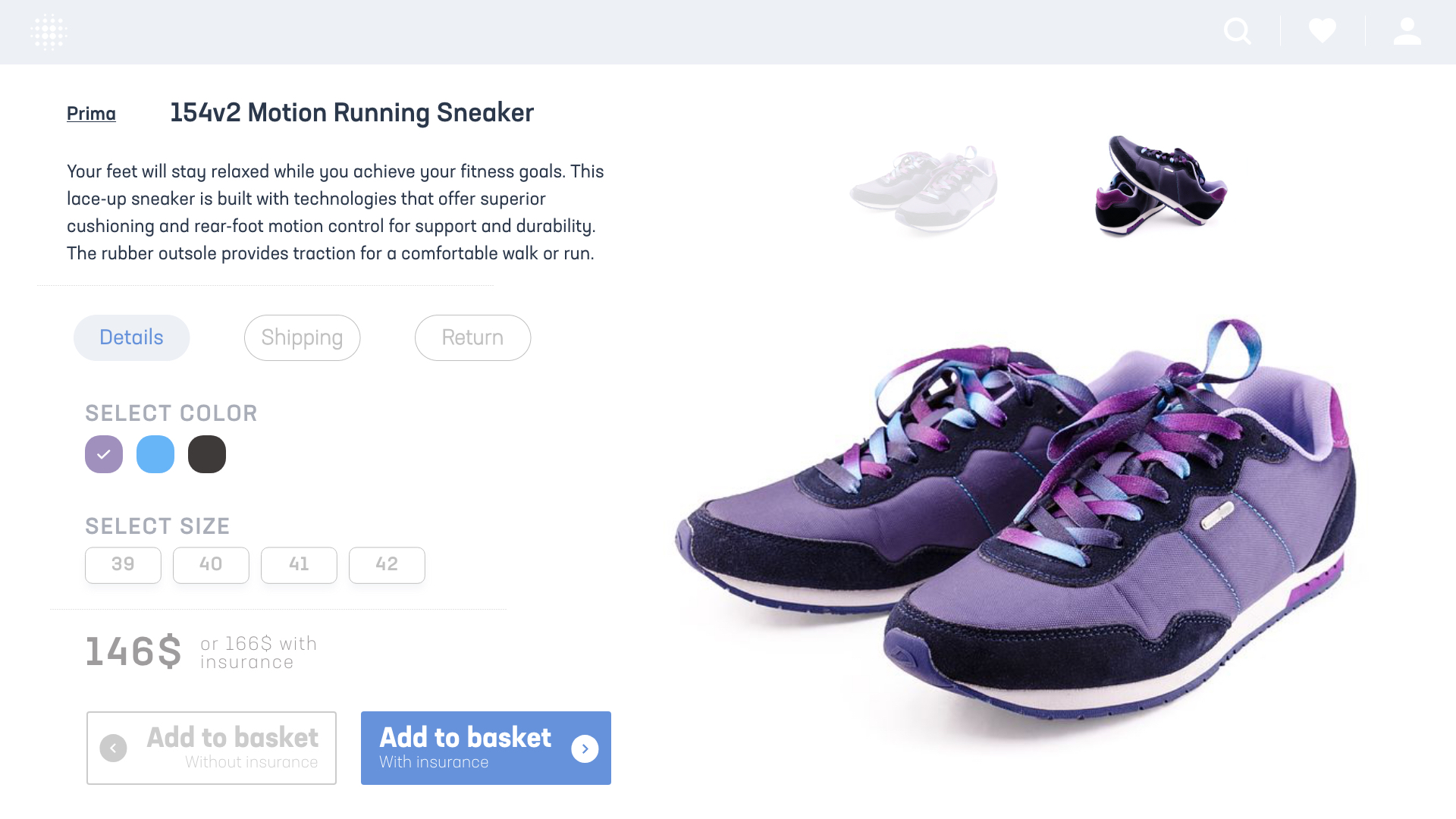}
        \label{fig:sneakbasket}
    \end{subfigure}
    \begin{subfigure}{0.49\textwidth}
        \includegraphics[width=0.9\linewidth]{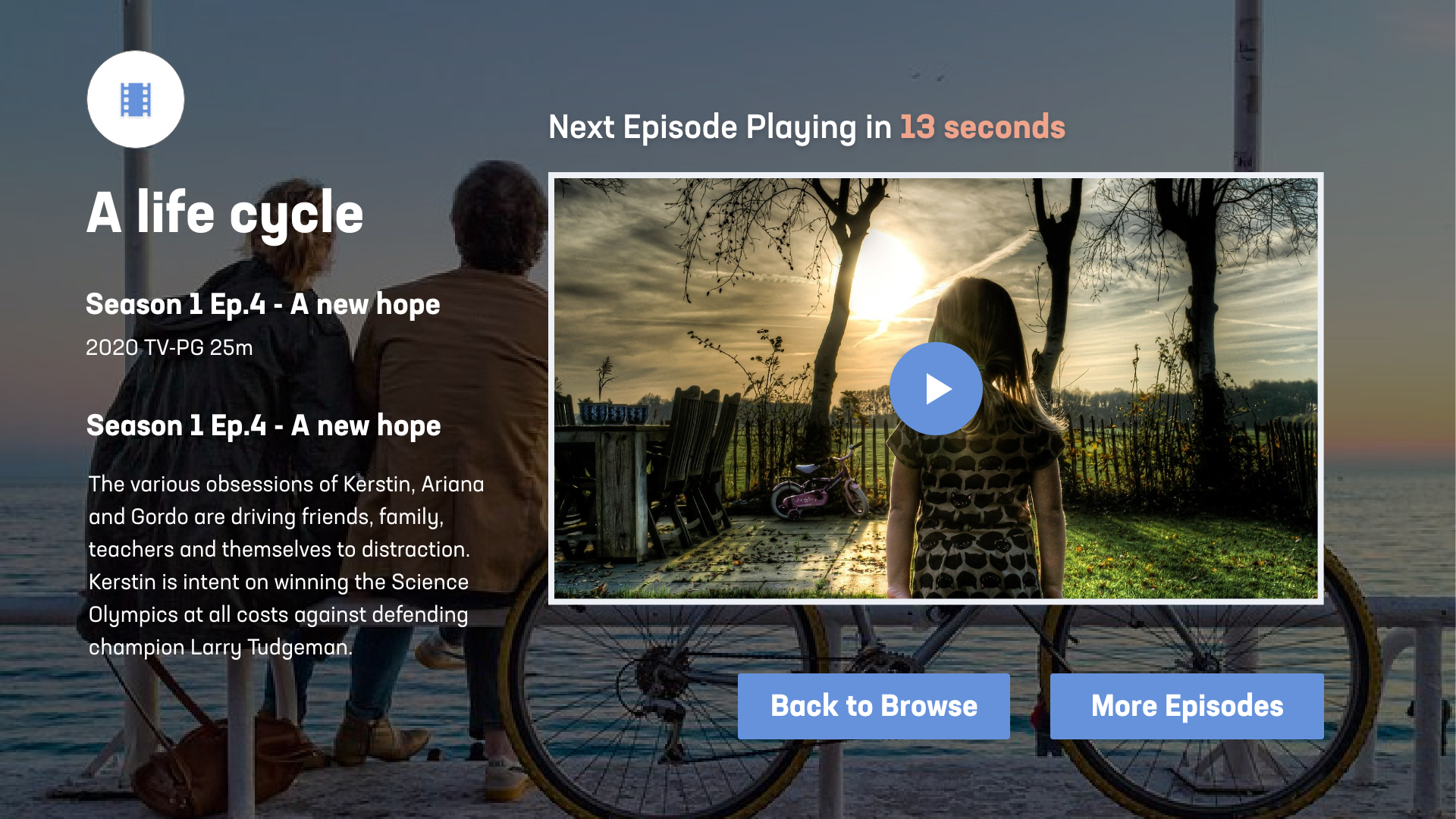}
        \label{fig:autoplay}
    \end{subfigure}
    \caption{Two example interfaces from the study. Left: (a) Sneak-basket / false hierarchy, right: (b) Auto-play}
    \label{fig:DPexamples}
\end{figure}

By testing nine different dark patterns (with consequences on finances, privacy, and time/attention, see two examples in Figure \ref{fig:DPexamples}) in the survey, we found that the scores for dark pattern detection, potential behavioural influence and acceptability vary among the patterns (Table \ref{table:DP data}). For example, both \textit{(a) sneak-basket / false hierarchy} and \textit{(b) auto-play} were recognised by about half of the participants. After receiving an explanation about what could be considered manipulative in the examples, the participants tended to find \textit{(b) auto-play} more influential than \textit{(a) sneak-basket / false hierarchy}. However, they deemed \textit{(b) auto-play} more acceptable than \textit{(a) sneak-basket / false hierarchy}. This implies that the \textbf{admissibility of a certain design is not necessarily related to its ability to influence}.

\begin{table}[ht]
\begin{tabular}{p{5.7cm}p{0.8cm}p{0.8cm}p{0.8cm}p{1.9cm}p{1.9cm}}
& \multicolumn{3}{l}{Dark pattern detected} & Influential & Acceptable\\ 
Dark pattern name & \textit{no} & \textit{partly} & \textit{yes} & &\\
 \hline
a) Sneak-basket / false hierarchy & 42\% & 4\% & \textbf{54}\% & -0.49(SD1.28) & -1.00(SD1.09)\\
b) Auto-play & 40\% & 11\% & \textbf{49}\% & 0.72(SD1.05) & 0.74(SD0.87) \\ 
c) Trick question / Preselection & \textbf{51}\% & 9\% & 40\% & -0.14(SD1.22) & -0.73(SD1.12)\\
d) Framing / Confirmshaming & 27\% & 19\% & \textbf{53}\% & -0.03(SD1.26) & -0.79(SD1.06)\\
e) Preselection / Framing & \textbf{51}\% & 11\% & 38\% & 0.00(SD1.20) & -0.50(SD1.08)\\
f) Hidden information / Trick question & \textbf{64}\% & 20\% & 16\% & 0.36(SD1.21) & -1.17(SD1.06)\\
g) Bundled/forced consent & \textbf{50}\% & 36\% & 14\% & 0.36(SD1.15) & -0.64(SD1.06)\\
h) High-demand / Limited-time message & 5\% & 11\% & \textbf{84}\% & 0.39(SD1.25) & -0.39(SD1.11)\\
i) Confirmshaming & 27\% & 2\% & \textbf{71}\% & -0.85(SD1.15) & -0.22(SD1.10)\\
 \hline
\end{tabular}
\caption{Dark pattern detection percentages for nine tested interfaces, and scores of participants' auto-evaluation of their likeliness to be influenced and their opinion on the acceptability of the strategy (from -2 = \textit{strongly disagree} to 2 = \textit{strongly agree})}
\label{table:DP data}
\vspace{-7mm}
\end{table}

\paragraph{Defining the problem} The spectrum of possible dark pattern design implementations is vast. There is not one single intervention that could free the web from all dark patterns. \textbf{Drafting appropriate interventions is hence in itself a design problem}. Before working on solutions, it is thus indispensable to understand which issue the intervention aim to solve. For example, certain dark patterns are not easily recognisable. In this case, depending on the pattern's attributes, interventions would have to either hone people's skills to recognise the pattern or eliminate the pattern from online interfaces all-together. For patterns that users recognize but find hard to resist, interventions that support people's self-protection efforts are required. One should also consider the degree to which users deem certain patterns tolerable (e.g., because there is a counter-weighting benefit). While interventions that counteract unacceptable patterns should strive to eliminate the pattern, raising user awareness might be more appropriate for patterns that people find acceptable.
Interventions need to be targeted and can come as a combination of educational, technical, design and regulatory measures. In the following section, we propose a matrix of interventions spaces and suggest measures accordingly. However, the discussion about different interventions would strongly benefit from more empirical research that links dark pattern attributes and the user perspective.\label{sec:1}

\section{Towards targeted interventions}
\subsection{Intervention spaces} 
\begin{figure}[hbt!]
    \centering
    \includegraphics[scale=0.13]{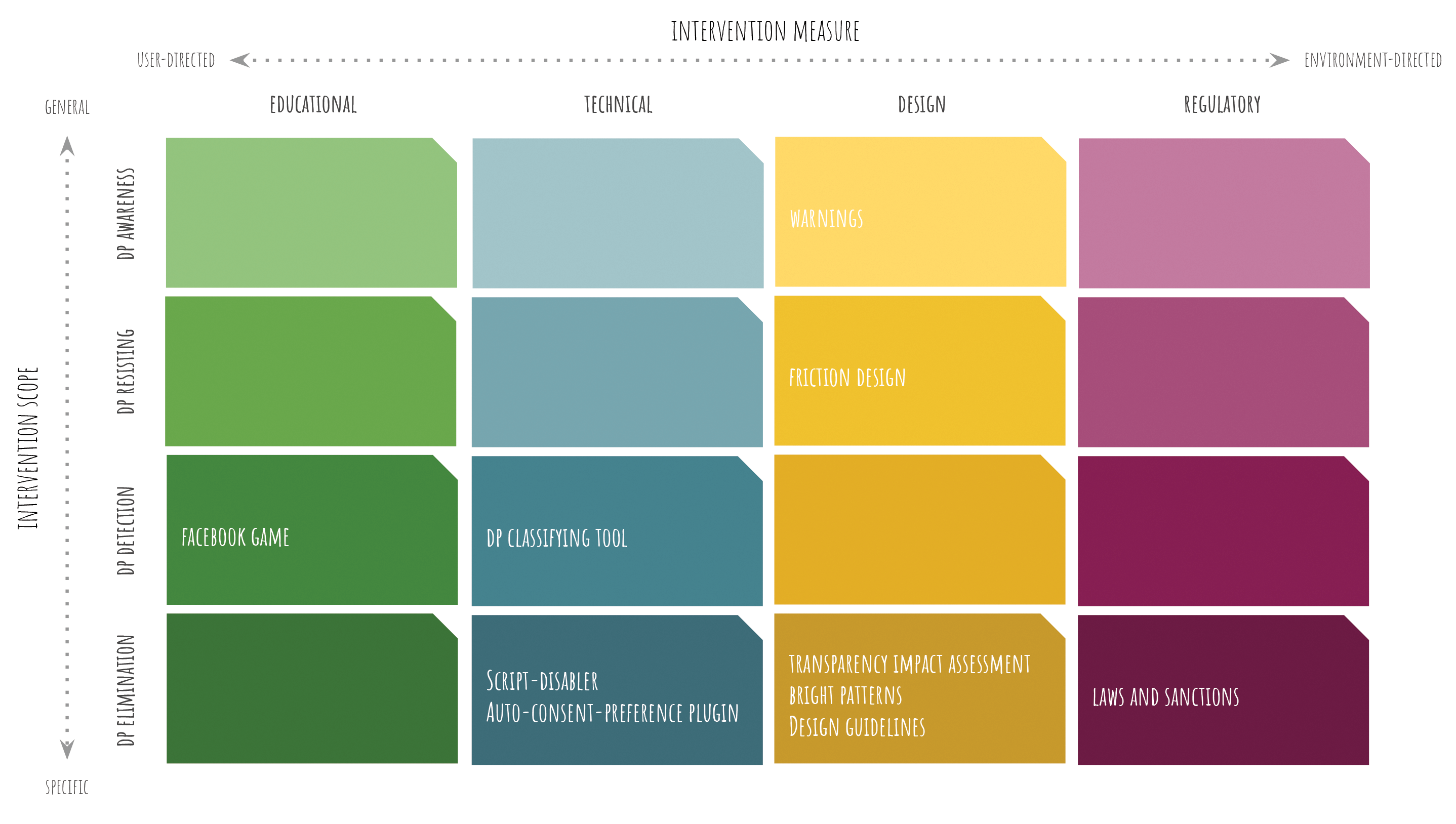}
    \caption{Intervention spaces for counteracting dark patterns (DP)}
    \label{fig:intervention spaces}
\end{figure}

We propose considering the interventions' scope and measure, to define both the actors that should support them and appropriate indicators to assess their success. Interventions can aim to (i.e., \textbf{scope}): a) raise awareness of the existence and the risks of dark patterns, b) bolster resistance towards them, c) facilitate their detection, or d) eliminate them from online interfaces. Interventions can act on the user or the environment (i.e., \textbf{measures}): while educational interventions favour users' agency, regulatory interventions tend to protect the user, with technical and design interventions situated in-between. The resulting matrix is shown in Figure \ref{fig:intervention spaces}. Certain spaces are left blank to stimulate the discussion during the workshop\footnote{\url{https://darkpatternsindesign.com/}.}.

\subsection{Educational measures}
Designers and researchers can explore \textbf{manipulation-protection strategies} with target users, like cause-and-effect data privacy scenarios and procedural rules \cite{grassl2020dark} (e.g., every time I encounter a cookie consent request, I look for the "refuse all" button). \textbf{Folk models} \cite{wash2010folkmodel} that trigger users' scepticism towards certain interfaces (e.g., sudden interruptions \cite{Bhoot2020IndiaCHI}) should also be further researched. \textbf{Gamified experiences} integrated into major digital services (e.g., a Facebook game) could strengthen the motivation to learn how to counter dark patterns in real settings, without the cognitive cost of transferring skills learnt in a different (e.g., offline, professional) context. 

\subsection{Technical measures}
\textbf{Tools that  automatically identify, flag, and even classify potential dark patterns} at large scale can be developed \cite{mathur2019dark,nouwens2020dark} to warn users, to expedite watchdogs' supervising tasks and to provide proofs of unlawful influence to consumer advocates. As such tools need large pools of reliable data, we are creating a database of examples of dark patterns published on Reddit, Twitter and Tumblr. Technical interventions can also ease autonomous decision-making, like the \textbf{add-on extension} Consent-O-Matic\footnote{\url{https://addons.mozilla.org/en-US/firefox/addon/consent-o-matic/}.} or  \textbf{plugins that disable (e.g., WordPress) plugins} that  create scarcity messages.

\subsection{Design measures}
\textbf{Bright patterns}, i.e., digital nudges that counteract dark nudging strategies, modify the environment where users make choices (e.g., assign the same visual weight to consent options vs pre-selections). A rich literature describes nudges that enhance privacy choices \cite{acquisti2017nudges} and that can even be personalised based on the individual's decision-making style to maximise their effectiveness \cite{Peer2020nudgeright}. However, dark patterns that use coercion and deception (e.g., deceptive framing, forced consent) cannot be simply counterbalanced by digital nudges. \textbf{Friction designs} are now widespread on streaming services (e.g., YouTube, Netflix) to counter binge-watching. Similar nudges could oppose not only infinite scrolling and defaults, but also mindless consent to data sharing. \textbf{Warnings} \cite{grassl2020dark} can make dangers salient and concrete (e.g., about financial losses) and counterbalance the human tendency to underestimate online threats because of, e.g., hyperbolic discounting and optimism bias. \textbf{Reframing the costs} of falling prey to dark patterns in terms that are personally relevant \cite{Moser2019impulse} may also be considered. For instance, by converting the time spent on binge-watching or scrolling into other pleasurable activities (e.g., 1 hour of Tik Tok swiping = 1 hour of reading).
Dark patterns can also be offset by official \textbf{design guidelines} \cite{CNIL_donnees,acm2020guidelines,age_designcode} and companies' involvement in problem-solving activities on concrete case studies \cite{CNIL_donnees}. Designers and product owners can employ ethical design tool-kits\footnote{E.g., \url{https://ethicalos.org}.} to envisage the consequences of their design choices. Simultaneously, persuasive technology heuristics (e.g., \cite{Kientz2010heuristic}) can be adapted to assess digital products' potential manipulative effects even before release. We plan to develop a \textbf{standardised transparency assessment tool} for digital services based on \cite{Meske2020ethical,Mathur2021what}.

\subsection{Regulatory measures}\label{sec:regulation}
Given the pervasiveness of dark patterns \cite{mathur2019dark,di2020ui}, business' scarce aptitude to self-regulation and human bounded rationality, \textbf{legal safeguards} should apply more stringently, as many dark patterns are unlawful, at least under EU consumer law \cite{acm2020guidelines} and the data protection regime \cite{EDPB_consent2020}. Hence, services employing dark patterns risk \textbf{stiff penalties}: in France, Google received fines for a total of 150 million € due to invalid consent design and elicitation \cite{CNIL2019google,CNIL2020google}. Empirical research demonstrating the presence, diffusion and effects of manipulative designs might have an impact on legal enforcement: cookie consent dialogues increasingly offer privacy-by-default options as a result of case law (e.g.,  the landmark case Planet49) and, conceivably, of intense academic scrutiny. The threat of more stringent regulations (e.g., the US' Social Media Addiction Reduction Technology Act) might be fuelled by public pressure (e.g., derived from the popularity of documentaries like ``The social dilemma'') and encourage self-regulation.\label{sec:2}

\section{Future work}
In this position paper, we have proposed a space to frame possible interventions and target them to the specific problem(s) we aim to solve. We suggest exploring established \textbf{dark pattern attributes in combination with the user perspective}: only by understanding to what extent certain (combinations of) attributes are commonly perceived as unrecognisable, irresistible and/or unacceptable by the users, can we devise appropriate interventions. Moreover, the exploration of user perception can help establish what is deemed legitimate without taking a normative stance. The intervention space may be enriched with additional dimensions, like measures that assess the interventions' effectiveness or economic measures.\label{sec:conc}

\section*{Acknowledgement}
We consider this publication as the first step of the project Decepticon (grant no. IS/14717072) supported by the Luxembourg National Research Fund (FNR).

\bibliographystyle{ACM-Reference-Format}
\bibliography{references}

\end{document}